# The Lindy Effect

Toby Ord  (University of Oxford)

The Lindy effect is a statistical tendency for things with longer pasts behind them to have longer futures ahead. It has been experimentally confirmed to apply to some categories, but not others, raising questions about when it is applicable and why. I shed some light on these questions by examining the mathematical properties required for the effect and generating mechanisms that can produce them. While the Lindy effect is often thought to require a declining hazard rate, I show that it arises very naturally even in cases with constant (or *increasing*) hazard rates — so long as there is a probability distribution over the size of that rate. One implication is that even things which are becoming less robust over time can display the Lindy effect.

One book has been in print for three years; another for three hundred. Which should we expect to go out of print first?

The *Lindy effect* is a statistical regularity where for many kinds of entity: the longer they have been around so far, the longer they are likely to last. This was first clearly posed by Benoît Mandelbrot (1982, p. 342) in his book, *The Fractal Geometry of Nature*:

> 'However long a person's past collected works, it will on the average continue for an equal additional amount. When it eventually stops, it breaks off at precisely half of its promise.'

Mandelbrot called this effect 'Lindy's Law' in honour of an anecdote by Albert Goldman (1964) about how the future career length of a comedian might be predicted from their past exposure.[1]

The idea was developed by Nassim Taleb (2012) in his book, *Antifragile*. The book focused on a special category of entities: those which aren't weakened by exposure to shocks and stresses, but which instead become stronger and more robust. He describes the Lindy effect in those terms (p. 318):

---

[1] It is not clear whether Goldman, or the denizens of Lindy's Deli he describes, were really describing the same principle as Mandelbrot.

> 'If a book has been in print for forty years, I can expect it to be in print for another forty years. But, and that is the main difference, if it survives another decade, then it will be expected to be in print another fifty years. This, simply, as a rule, tells you why things that have been around for a long time are not "aging" like persons, but "aging" in reverse. Every year that passes without extinction doubles the additional life expectancy. This is an indicator of some robustness. The robustness of an item is proportional to its life!'

Taleb (pp. 317–18) suggests that this Lindy effect applies to all things without a natural hard upper bound to their lifespans (those that are 'non-perishable'[2]):

> 'For the perishable, every additional day in its life translates into a *shorter* additional life expectancy. For the nonperishable, every additional day may imply a *longer* life expectancy.'

The Lindy effect is an important and much-discussed principle but has had surprisingly little formal development. This paper attempts to address that need. In doing so, it takes a synoptic view: connecting the Lindy effect to results and techniques from statistics, reliability analysis, economics, population ecology, and cosmology.

**Which distributions produce which Lindy effect?**

The Lindy effect doesn't yet have an agreed precise definition.[3] In some cases it is suggested to be the narrow claim that something's future life expectancy is exactly equal to its age (e.g. Mandelbrot 1982). In others, it is the broad claim that future life expectancy increases with age.[4] And sometimes it is said to be the *median* future lifespan (the estimate which minimises the absolute prediction error) that equals the current age (e.g. Eliazar 2017).

---

[2] He further suggests that to be non-perishable an item must have an informational component, but this appears to be neither necessary nor sufficient. Many physical entities that do not display the Lindy effect (such as humans) are partly informational (with a genetic code that allows our internal systems to actively repair damage), and some physical entities that do display the Lindy effect (such as buildings) do not seem to be any more informational.

[3] One could even define a version of the Lindy effect that did not refer to lifespans, since a very similar phenomenon can occur directly in other quantities that accrue over time (such as the number of people who have bought a particular album by now versus the number who will buy it in the future). Indeed, sometimes the generating mechanisms for a Lindy effect can apply directly to those quantities rather than to time (e.g., you might reasonably expect your purchasing of the album to be uniformly sampled from all purchases that will be made). A version of the Lindy effect could even apply to quantities that have no connection to time, so long as one of the generating mechanisms is present. For example, if you pick a random house on a street, and find it is numbered $n$, Gott's generating mechanism suggests your posterior for the number of houses on the street is a power law with median $2n$ (demonstrating a connection between the Lindy effect and the German tank problem).

Any of these kinds of Lindy effect can be understood in terms of the probability distribution of the total lifespan *T* and how this depends on the age so far, $t_0$. Distributions are usually analysed in terms of their PDF, *f*(*t*), or their CDF, *F*(*t*). But in studying the Lindy effect it is most convenient to adopt the methods of survival analysis (Miller 2011) and think in terms of the distribution's survival function, *S*(*t*), and hazard function, λ(*t*).

Any one of these four functions completely determines the probability distribution, but each foregrounds a different feature. The survival function represents the probability that the entity in question is still alive at time *t*: *S*(*t*) = Pr(*T* > *t*) = 1 − *F*(*t*). The hazard function is the probability density of the lifespan ending at time *t*, conditional on having made it that far already: λ(t) = *f*(*t*)/ *S*(*t*). The two are related by the fact that survival decays exponentially in the area under the hazard curve up to that time:

$$S(t) = e^{-\int_0^t \lambda(t')dt'}$$

For the special case of a constant hazard rate, λ(*t*) = *k*, survival decays exponentially: *S*(*t*) = $e^{-kt}$. This exponential distribution has the special property of being *memoryless* — the time remaining before an event is entirely independent of how much time has already elapsed. So something whose lifespan is exponentially distributed (such as a radioactive particle) does not display the Lindy effect.

But if the hazard rate were to decline monotonically over time, then the survival curve would decline more slowly than an exponential (such distributions are known as *heavy-tailed*). This implies that the time needed for survival to halve would keep getting longer the more time that has elapsed — these distributions thus have medians and means for the remaining time before the event (*T* − $t_0$) that grow when the elapsed time so far ($t_0$) grows. Therefore, the exponential distribution (and equivalently the constant hazard rate) marks the border line for the most general kind of Lindy effect: all heavy-tailed distributions (and declining hazard rates) exhibit the Lindy effect in the broad sense.

For the narrower definitions, there is a close connection between the Lindy effect and power law distributions (a subset of heavy-tailed distributions). Eliazar (2017) has shown that if a narrow Lindy effect holds across the entire domain of possible

---

[4] Taleb (2012, p. 318) clarifies in a footnote that for him the situation where future life-expectancy exactly equals age is a special case. Officially he takes the Lindy effect to be the broad claim that 'the non-perishable has a life-expectancy that *increases* with every day it survives.' [The italics appear in the original.]

lifespans — such that the expected future lifespan is always equal to $p$ times the past lifespan — then the total lifespan, $T$, must obey a Pareto distribution with a survival function of:

$$S(t) = Pr(T > t) = \begin{cases} (\frac{t_0}{t})^\alpha & , t \geq t_0 \\ 1 & , t < t_0 \end{cases}$$

Where $\alpha = 1 + 1/p$. In the case specified by Mandelbrot and Taleb where $p = 1$, the exponent is equal to 2, and thus the expected future duration is power law distributed with an inverse square tail on the survival function. This is a distribution with mean of $2t_0$ and infinite variance (see *Figure 1*).

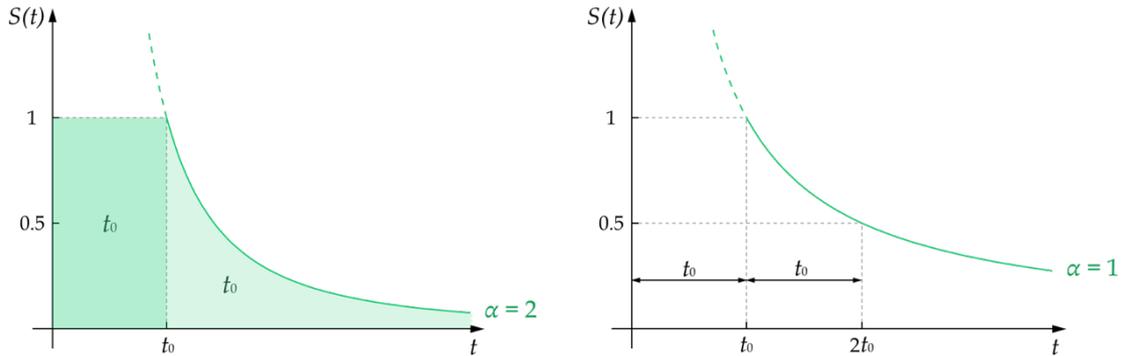

*Figure 1. Key power law survival curves.* LEFT: The survival curve for a Pareto distribution with $\alpha = 2$. The area of the green rectangle leading up to any point in time is always equal to the area under the curve from that point on. These areas correspond to the lifespan so far and the expected future lifespan. RIGHT: The survival curve for a Pareto distribution with $\alpha = 1$. For any starting point, it will always take that long again before the chance of surviving halves.

What if one instead requires that the *median* future lifespan is always equal to $q$ times the past lifespan? The total lifespan, $T$, is again Pareto distributed but with $\alpha = \ln(2) / \ln(1+q)$. In the special case where the median future lifespan is always equal to the past lifespan, $\alpha = 1$. This gives a survival function with a hyperbolic tail: $S(t) = t_0/t$. This has the elegant property that the chance of surviving beyond $3t_0$ is $1/3$, beyond $4t_0$ is $1/4$, and so on. However, it is an extreme distribution that is not always easy to work with, since both its mean and variance are infinite.

Thus, several natural ways of narrowly defining a Lindy effect correspond to lifespans being distributed through a power law — in particular, a Pareto distribution. But note that while there is a convergence in form, there is also something of a clash in specifics. Only one out of the expected future lifespan or the median future lifespan can be always equal to the lifespan so far, so these definitions are incompatible, and proponents of a narrow Lindy effect need to work out which one it is (if either) that is supposed to govern so many real-world phenomena.

While each definition corresponds to a range of different Pareto distributions (for different values of $p$ or $q$), the formulation in terms of median is more expressive (Eliazar 2017). It allows Pareto distributions with any exponent, $\alpha > 0$, to qualify as the underlying distribution for a Lindy effect. The formulation in terms of expectations rules out power law distributions where α ≤ 1 because the expected future lifespan is so great as to already be infinite and thus can't be shown to increase when $t_0$ increases.

**A mechanism for the Lindy effect**

In the early 19th century, Pierre-Simon Laplace attempted to use probability theory to answer a challenging (and somewhat ill-defined) question about predicting the probability of something coming to an end imminently, based purely in terms of how long it has been going so far — for example the probability that the sun fails to rise tomorrow conditional on it having risen every one of the $n$ days so far. The sample mean ($0/n$) would give a probability of failure of precisely zero, but presumably we can't be absolutely certain. So we might instead think this probability should be non-zero, yet decreasing as our evidence builds up with more and more trials without a failure.[5]

Laplace (1814) suggested we consider an urn containing an unknown quantity of black and white balls. We repeatedly draw balls at random from the urn (replacing them after each draw) and our first $n$ attempts all retrieved white balls. What is the probability that the next ball is black? Laplace modelled this as a Bayesian inference problem where you begin with a uniform prior over [0, 1] for the fraction of black balls, then update this with each successive draw. The resulting formula is known as Laplace's Law of Succession and gives a probability of $1/_{n+2}$ for the chance that the run of white balls comes to an end on the next draw.

We can also use this model to ask how long this run of white balls will continue, based on its length so far. The answer is a (discrete) power law distribution with α = 1, and thus a median of $n$ more draws before a black ball and an infinite mean. This is an exact discrete analogue of the median formulation of the narrow Lindy effect.

This convergence between Laplace's Law of Succession and the Lindy effect is intriguing. Not only is it another pointer towards the formulation in terms of the median, but it also manages to derive a Lindy effect out of repeated draws from an urn, which have a constant probability of the streak coming to an end. In other words, there is a Lindy effect despite a constant hazard rate — something we seemed to have previously ruled out.

What is happening is that in this case there are two nested probability distributions at play: one for the unknown fraction of black balls and one for the number of draws from the urn before drawing a black ball. The former is a uniform distribution and the latter an exponential distribution, but combined they make a power law distribution exhibiting the narrow Lindy effect.

---

[5] This general problem is still studied within reliability analysis under the banner of 'the problem of zero-failure data' (Bailey, 1997; Quigley & Revie, 2011).

Can this also happen in continuous time? Let's suppose, like Laplace, that there is a constant but unknown hazard rate *k* representing a physical probability, like that of a coin toss or (in continuous time) a radioactive particle decaying. And following Laplace, let's suppose that our prior for *k* is uniformly distributed over some interval from 0 to *K*. In our example this could be our uncertainty about that isotope's rate of decay. We know that ultimately *S*(*t*) will be an exponential decay curve, but we don't know which one. There is a kind of superposition of different exponential decay curves, $S_k(t)$, that the particle could be on.

In this case, we can calculate an effective survival curve combining our epistemic uncertainty about the parameter *k* with the physical chance of surviving for some duration given that parameter value (see *Figure 2*). Each $S_k(t)$ can be thought of as the overall probability of surviving beyond time *t* given the value of *k*: i.e. Pr(*T* > *t* | *k*). We want to know the unconditioned probability.

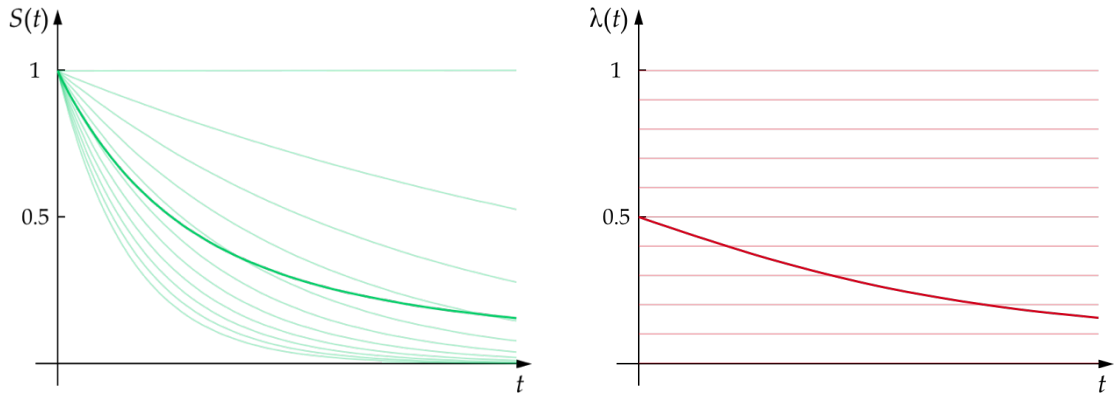

*Figure 2. Superposition of Survival Curves.* LEFT: The faint green curves are exponential survival curves with hazard rates uniformly spread between 0 and *K*. If we assume equal chances of any of those faint curves being the true survival curve, then the bold green curve is the effective survival curve (the chance of surviving until that point averaged across all the survival curves). RIGHT: The faint red curves represent the constant hazard rates corresponding to each of the possible survival curves. The bold red curve is the effective hazard curve (determined via the effective survival curve). Its decline over time indicates a Lindy effect.

We can find this by computing the average probability of surviving until *t* according to all different values of *k*, weighted by the probability that *k* takes on that value (i.e. the expected survival at any time *t*):

$$S(t) = \mathbb{E} S_k(t) = \int_0^K S_k(t) \rho(k) dk = \frac{1}{K} \int_0^K e^{-kt} dk = \frac{1 - e^{-Kt}}{Kt}$$

As *K* or *t* get larger, the term $e^{-Kt}$ quickly becomes trivial, leaving:

$$S(t) \approx \frac{1}{Kt}$$

So for any uniform prior for the hazard rate (up to a specified maximum: *K*), *S*(*t*) approaches a power law tail with exponent α = 1. This is just what was needed for the long-term behaviour to approach a narrow Lindy effect where median future lifespan equals past lifespan.

We can also think of all this in terms of the effective hazard function that corresponds to this effective survival curve (see *Figure 2*). While each of the possible physical hazard curves is constant, the effective hazard curve is declining, producing a Lindy effect.

We can determine the effective hazard from the effective survival by the standard formula:

$$\lambda(t) = -\frac{d}{dx}S(t)$$

And so in the limit of high *Kt*:

$$\lambda(t) \approx -\frac{d}{dx}\frac{1}{Kt} = \frac{1}{Kt^2}$$

How can effective hazard be declining? While each of the hazard curves is constant, their relative weighting is changing. We can interpret this in Bayesian terms: the longer you survive, the more evidence you accumulate against higher hazard rates.[6] Since the probability of observing survival to time *t* at hazard rate *k* equals $S_k(t)$, the Bayesian update to the uniform distribution of hazard rates weights each rate *k* by $S_k(t)$, giving decreasing weight to the higher hazard rates and thus a declining effective hazard rate.

So it is possible to have a Lindy effect even when the object itself is not becoming more robust over time. For example, the physical properties of a radioisotope (of unknown half-life) are not changing as time proceeds — it remains equally robust — though your expectation of its future lifespan keeps increasing. Its *observed* or *estimated* robustness is increasing, but this is because you are gaining information that it is more likely to be a robust specimen if it has made it this far. There is an *evidential* increase in its robustness as time passes without failure, but no *causal* increase. The change is in the observer, not in the object being observed.

---

[6] This interpretation also helps us see how this Bayesian Lindy effect combines with other forms of evidence about the distribution. For example, after witnessing a life table for human mortality, your distribution over the size and shape of the hazard function becomes sufficiently narrow (aggregating the observed lifespans of millions of people) that the one additional data point of someone's survival up to time $t_0$ updates it very little.

The decline in the hazard rate can also be understood in frequentist terms. If the population had a mixture of members with each different constant hazard rate *k*, then over time a higher share of members with higher hazard rate will not survive to that time, biasing the mixture towards members with lower hazard rates. In particular, the share of members with hazard rate *k* will be proportional to their surviving fraction $S_k(t)$, giving the same mathematical result.

This fits a general observation in survival analysis. When objects have increasing hazard rates, this is generally due to cumulative damage or wear, which is sometimes called 'aging'. When they have declining hazard rates this can be due to something like what Taleb calls 'aging in reverse' or 'antifragility', where there is a cumulative process that increases inherent robustness, such as an organism or institution learning from and adapting to the stresses so far. But it can also be a kind of winnowing process, where entities with defects fail quickly, leaving behind a greater proportion of those which were more robust all along.

And indeed, this latter explanation is sometimes mentioned in explanations of the Lindy effect.[7] A book that has been in print longer has survived a market test, demonstrating its enduring interest. A building that has stood longer has passed the test of time, demonstrating its superior materials (or its enduring desirability to those who might maintain it).

Can this winnowing mechanism for the Lindy effect be generalised further? What if we relaxed the condition on the hazard rate, so that it no longer needed to be constant? Let's allow it to grow (or decline) over time as a power law: $\lambda(t) = kt^\beta$ for some specified exponent β but unknown *k* — again distributed uniformly between 0 and *K*.[8]

$$S(t) = \frac{1}{K}\int_0^K e^{-\int_0^t \lambda_k(t')dt'}dk = \frac{1}{K}\int_0^K e^{-\int_0^t kt'^\beta dt'}dk$$

Solving the integral gives:

$$= \frac{(\beta+1) - e^{-Kt^{(\beta+1)}/(\beta+1)}}{Kt^{(\beta+1)}}$$

Again, as *K* or *t* grows, the exponential term in the numerator quickly becomes insignificant, leaving us with the long-run behaviour for *S*(*t*):

$$\approx \frac{\beta+1}{Kt^{(\beta+1)}}$$

---

[7] For example, Taleb (2012, p 319) remarks, 'I am not saying that *all* technologies do not age, only that those technologies that were prone to aging are already dead.'

[8] This flexible distribution with the hazard rate changing as some power of time is closely related to the Weibull distribution.

And this is again a power law tail. This time α = β + 1. So even if an object had an intrinsic hazard rate that increased as a high power of time, uniform uncertainty about the scaling parameter on that hazard rate would create an overall distribution with a power law tail and would thus produce a broad Lindy effect.

But this is as far as it goes. If we push the rate of growth of the hazard to exponential, or beyond, the expected survival curve is no longer heavy-tailed and the Lindy effect disappears:

$$S(t) = \frac{1}{K} \int_0^K e^{-\int_0^t ke^{t'} dt'} dk = \frac{1 - e^{K(1-e^t)}}{K(e^t - 1)} \approx \frac{1}{Ke^t}$$

This may help to explain why the empirical distribution of human lifespans doesn't exhibit even a broad Lindy effect: in the latter years of our lives, the hazard rate rises exponentially. It thus may not be that humans are crisply in a category of 'perishable' things with hard upper bounds on our lifespans, but rather than we are in the more subtle category of things with exponentially increasing hazard.

Looking in the other direction, one can consider hazard functions that *decline* over time as a power law $\lambda(t) = kt^\beta$, for negative β. So long as β > –1, the earlier approximation still applies and this still gives a Lindy effect.[9] This is not too surprising as we had seen that declining power law hazard functions give a Lindy effect, even without any uncertainty, but it is useful for showing the robustness of the phenomenon: all power law hazard functions with uniform uncertainty over their scale produce versions of the Lindy effect.

How robust is this effect to the choice of distribution for the size of the hazard? One key feature is that for the power law (and thus Lindy effect) to hold in the long term, it is essential for $k$ to be able to be arbitrarily close to zero. It doesn't matter if the interval for the range of $k$ is open at zero, or (in some cases) if the probability density reaches zero exactly at $k = 0$, but if it is certain that $k > \epsilon$, for some specified $\epsilon > 0$, then the tail of the effective survival function eventually approaches an exponential and the Lindy effect fades away as time goes on.

---

[9] What about β < –1? A simple answer is that hazard *can't* decrease that quickly. In order to generate a PDF that integrates to probability 1, cumulative hazard needs to rise without bound — this is part of the standard definition of a hazard rate. If cumulative hazard were to remain bounded, the object would have a nonzero chance of lasting forever: $\Pr(T = +\infty) > 0$. This is now a distribution over the extended reals, and one could extend this model to cover it. We can no longer track changes in mean or median $T$ (as both become infinite), but we can see that an object's chance of lasting forever is increasing in $t_0$ — a generalisation of the broad Lindy effect.

**Alternative generating mechanisms**

In 1993, Richard Gott presented a remarkable series of observations about predicting future lifespans by applying the Copernican principle.[10] This is a generalisation of the idea that humanity should expect to find itself on a typical planet, rather than a special one. The Copernican principle suggests that unless there is evidence to the contrary, we should expect to find ourselves in typical situations rather than special ones, and the principle is frequently deployed in astronomy and cosmology. Gott's argument achieved notoriety (and the name 'The Doomsday argument') when he applied it to estimating the future lifespan of *Homo sapiens*, but he has also applied it (with considerable success) to more mundane topics like the future lifespans of Broadway plays and the future reigns of world leaders (Tyson et al. 2016, pp. 436–7).

One of Gott's observations (Gott 1994) can be thought of as providing a Bayesian case for a very general kind of Lindy effect. He states that if we start with maximal ignorance about the lifespan of some entity, then we should begin with a 'vague prior' that assigns equal likelihood to a lifespan of any order of magnitude.[11] This is the maximum entropy prior for a variable that takes a positive real value, which is a prominent way of formalising having maximal ignorance. This log-uniform prior implies that the probability density for $T$ scales as $\rho(t) \propto 1/t$.

On observing a current lifespan of $t_0$, we update this prior for $T$ by re-weighting each hypothetical value of $T$ by how probable it would be to witness a current lifespan of $t_0$ according to that hypothesis. Our observation is incompatible with all values of $T < t_0$, so their probability is updated to zero. Gott suggests that given maximal ignorance, higher values of $T$ would make it less likely to witness a current age of $t_0$, since the number of other possible times we could have witnessed grows in proportion to $T$. He thus concludes that we must re-weight each possible value $t$ that $T$ could have by a factor of $1/t$. This gives $\rho(t) = 0$ for $t < t_0$, and $\rho(t) \propto 1/t^2$ thereafter (see *Figure 3*).

---

[10] He originally formulated the method much earlier, in 1969 on a visit to the Berlin Wall, when he used it to predict the wall would only last between 2.66 and 24 more years. When the wall fell 20 years later, he felt compelled to publish the method. A related argument was independently made by Brandon Carter (1974, 1983). Note that while Gott is often associated with the Lindy effect, he doesn't mention it explicitly and to the best of my knowledge, this is the first time the connections have been made mathematically explicit.

[11] This log-uniform prior is not a proper prior (it doesn't correspond to a probability distribution), but this need not be a problem in Bayesian reasoning as it becomes a valid probability distribution when conditioned on the evidence.

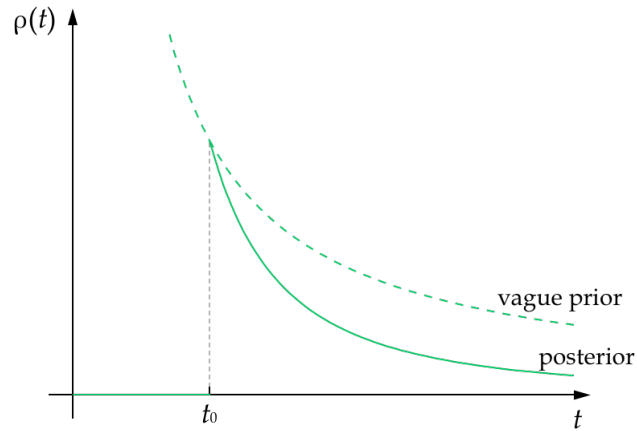

*Figure 3. Updating the vague prior to the Pareto distribution.* Starting with a vague (maximal entropy) prior, where ρ($t$) ∝ 1/$t$, we update on the fact that $T \geq t_0$ and end up with a Pareto distribution with α = 1.

This is precisely a Pareto distribution with α = 1 — exactly what is needed for a narrow Lindy effect where the median estimate of the remaining lifespan is given by the current age. One could think of this as an extreme version of the Laplacian idea that uncertainty can generate a Lindy effect. Here, a particular approach to dealing with maximal uncertainty in an objective Bayesian setting leads directly to a narrow Lindy effect for any type of entity at all.

In some sense, he has conjured a Lindy effect from nothing. Indeed, it is highly dependent on knowing exactly nothing — as soon as one knows *anything* more about the kind of entity or some further data about other lifespans of entities in that class, one must update the posterior distribution again and it will not generally remain a power law. Still, it is a remarkable argument for the Lindy effect and further work may be able to clarify how fragile the effect is when one incorporates more information.

Gott (1993) also provides an intriguing frequentist argument that can be adapted to support what we might call the 'Lindy estimate' — using $t_0$ as an estimate for the remaining lifespan. Consider the task of estimating how many more years a currently running Broadway play might last. Unless we have reason to suspect otherwise, we might treat the current moment as being sampled uniformly at random from all the times we could have asked this question — from the play's entire lifespan. If so, then there is a 95% chance that we are in the middle 95% of its lifespan (and similarly for any other interval). In other words, if it has been around for $t_0$ years so far and will have a full lifespan of $T$, then with 95% probability: $t_0 \in [\frac{1}{40} T, \frac{39}{40} T]$. This implies there is a 95% chance that $T \in [\frac{40}{39} t_0, 40 t_0]$.

Gott tested this empirically with the 44 Broadway plays that were running on the date his paper appeared and as of 2016, the 42 which had finished their runs all had lifespans within the estimated 95% confidence interval. Similarly, of the 313 world leaders in power on the date his paper appeared, 94% ended power within the given 95% confidence interval (Tyson et al. 2016, pp. 436–7).[12]

We can adapt this approach to generate a median estimate for when something will end. The method suggests a 50% chance that the current moment falls in the first half of the entity's total lifespan: that $t_0 \in [0, 0.5\, T]$, or equivalently, that $T \geq 2t_0$, and thus that $T - t_0 \geq t_0$. Similarly, consideration of the chance that the current moment is in the *second* half of the entity's total lifespan tells us there is also a 50% chance that $T \leq 2t_0$ and thus that $T - t_0 \leq t_0$. This Copernican principle therefore delivers a version of a median-based Lindy effect, where $t_0$ acts as a balanced estimate of remaining lifespan.

Puzzlingly, this argument can be applied even when we have full knowledge of a distribution and know that it has an increasing hazard rate. For example, consider the distribution of human lifespans. This distribution is known to not exhibit the Lindy effect — most of the distribution has an exponentially increasing hazard rate and beyond infancy the longer someone lives the lower their median and mean remaining lifespan. Yet Gott's arguments can be adapted to prove that even in this case if you estimate each person's remaining lifespan by their past lifespan, the estimate will be perfectly balanced, with a 50-50 chance of under-estimating or over-estimating it.[13]

We can resolve this apparent paradox via the following diagram (*Figure 4*). First consider each individual person in the population. Their lifespan can be represented as a horizontal bar stretching from zero to their age at death. Stacking these bars on top of each other recapitulates the survival curve.[14] We could now add vertical lines representing each different year of age. If we were to estimate each person's remaining lifespan once during each year of their life, we would have an estimate for every box in the diagram. If we use the narrow Lindy estimate of $t_0$ remaining years, then it always turns out that half of these estimates (those to the left of the blue line) would be under-estimates and the other half would be over-estimates.

---

[12] This included the heads of state and heads of government of independent countries. A few remained in power in 2016, so the final tally couldn't be completely confirmed, and Gott's 94% is on the assumption that none will remain in power past their 100th birthday.

[13] Gott (1996) attempted something very similar, but for condidence intervals rather than medians, showing that according to the world actuarial tables, 96% of people were expected to die within the 95% confidence intervals his method assigned.

[14] If the survival curve represents the empirical distribution of lifespans for the whole population, then the horizontal lines are evenly spaced as written. If it represents the idealised probability distribution, then they are instead distributed uniformly at random heights between zero and one. The main difference is that if it represents the idealised distribution, then the remaining claims are probabilistic instead of analytic. But with large populations, they occur very close to the analytic numbers with high probability.

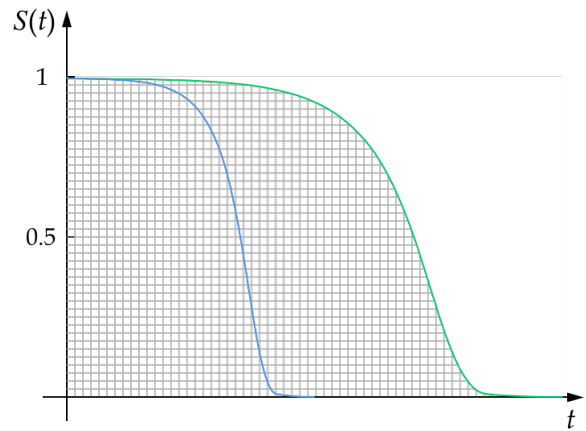

*Figure 4. Estimating each person's lifespan each year.* If we estimate each person's remaining lifespan every year of their life, we end up with an estimate for every box under the green survival curve. If we estimate the remaining lifespan as equal to their current age, our estimate has a perfectly calibrated chance of being too high or too low.

Thus there is an important sense in which the Lindy estimate of remaining lifespan is well-calibrated no matter what the empirical distribution of lifespan turns out to be.

But this is not sufficient to make it a good estimate. Given the empirical distribution of lifespans, one can calculate the true median remaining lifespan as follows. Start on the survival curve at their present age, then continue along it until the height of the curve has dropped by a factor of 2 — the horizontal distance between these points is the median remaining lifespan.

Like the Lindy estimate, the median estimate for their remaining lifespan is also well-calibrated. But it is also much more accurate. On a survival curve like this one (chosen because it doesn't exhibit the Lindy effect), when the Lindy estimate is wrong, it is usually very wrong. And being just as likely to be very wrong in either direction is not much comfort. In contrast, the median is the unique estimate that minimises the average error — the absolute number of years between the estimated lifespan remaining and the actual number. What was so special about the α = 1 power law (whether it arose directly in the survival curve, or due to uncertainty over the hazard rate), was that it is the unique case where the Lindy estimate equals the median estimate, such that it has accuracy as well as calibration.

Taleb (2021) has recently suggested another generating mechanism for the Lindy effect: Brownian motion with an absorbing barrier. The idea is that the entities whose lifespan we are estimating have some intrinsic parameter, $B(t)$, which we could think of as health or robustness. All entities start with the same level, $B(0)$, but it then fluctuates stochastically with time, following an unbiased continuous random walk (Brownian motion). If $B(t)$ ever falls below some minimal threshold, the entity's life ends. This mechanism generates power law survival times with exponent α = 1/2. This is a very heavy tailed power law, giving rise to a very strong Lindy effect, where the median lifespan remaining is always equal to 3 times the lifespan so far.

On this mechanism too, the Lindy effect wouldn't correspond to increasing robustness.[15] Instead, it would be a time-varying robustness which starts out the same for everything, but can rise or fall with equal probability. A longer past lifespan provides evidence to the observer that robustness is currently higher — and thus that future lifespan will be longer — but the longer past lifespan doesn't cause that. Rather than a case of anti-fragility, this would be a distinct way in which an uncertain parameter could produce a Lindy effect.

**Discussion**

We have seen that a single distribution for the lifetime of an entity produces a broad Lindy effect when it is heavy tailed and produces narrow Lindy effects for the mean and median when it is a Pareto distribution with α = 2 or 1, respectively. These are all distributions with declining hazard rates.

But if we have uncertainty about the scale of the hazard function, then it is also possible to get a Lindy effect when the hazard rate is constant, or even increasing.

This puts pressure on the idea that the real-world Lindy effect is generated by things being antifragile — becoming more robust in light of stresses they face. While our beliefs may still shift towards them being more and more robust as the stresses mount up, in some cases this does not reflect any improvements in the intrinsic nature of the entity itself. We may even *know* that the entity is getting more and more likely to fail over each successive year — it is just that this is being outweighed by our updates towards it being among the more robust objects of that type. Antifragile things can be Lindy, but so can fragile things.

While I presented this argument in terms of uncertainty about the magnitude of the hazard function, all that we really require is that there is a *distribution* over that magnitude. This distribution could come from uncertainty, but it could also come in the form of a population of objects, each with a different scaling of the hazard function, where the scalings are uniformly distributed between 0 and $K$. In this case, the population itself can be observed to display a decreasing hazard rate (and thus the Lindy effect) even when each member has a constant or increasing rate. We might worry that this is indistinguishable from a case where each object is intrinsically becoming less likely to fail, but if we had the ability to closely inspect each individual object to see its particular level of $k$, or had the ability to run repeated experiments on a single object, we could tell the difference.

---

[15] While it isn't clear if the parameter represents robustness directly, the parameter is the only thing that varies and lower values are associated with making the object fail sooner, so whatever robustness might mean exactly, it must be an increasing function of the parameter.

While we focused on the case of uniform distributions over the scaling of a power law hazard function, there are many shapes of distribution that could occur at each of the two levels, that would combine to produce a Lindy effect. At the extremes there are combinations where one level has a point-like distribution. For example, you could have a single kind of object whose lifespan is intrinsically Pareto distributed (the antifragility mechanism) or you could have a Pareto credence distribution over fixed lifespans that each object could have (Gott's Bayesian mechanism). Between these lie the intermediate cases, such as (roughly) uniform distributions over (roughly) constant hazard rates.

It is worth noticing that the forms of Lindy effect that we've discussed have come in two types. There is an *empirical Lindy effect*, where the effect appears in the observed distribution of lifespans, which can directly be shown to be heavy tailed. For the empirical Lindy effect, one can ask whether various entities have lifespans that display the effect and confirm the results experimentally. But there is also an *epistemic Lindy effect*, where it is our rational estimates for remaining lifespan that increase with observed lifespan, and thus our credence distribution over lifespans is heavy tailed. In this case, it isn't really open to experimental refutation. Instead, a better question would be to ask which combinations of priors, models, and evidence give rise to the effect.

Clearly these two forms are quite closely linked. Firstly, they obey the same mathematics. And they also interact with each other. Most notably, if you know with certainty the underlying distribution of lifespans for the entity in question, then there will be an empirical Lindy effect if and only if there is an epistemic Lindy effect.

We have seen that there are several close connections between the narrow Lindy effect and a Pareto distribution with $\alpha = 1$. One might worry that because this distribution has an infinite expectation it will necessarily be unphysical, and thus a bad model of the actual distribution of lifespans. However, this distribution is not like a model where an observable quantity (such as population) goes to infinity in finite time. The expectation of a distribution is not itself a physically realised quantity — just an idealised summary measure. And while this Pareto distribution does assume observed quantities could get arbitrarily large, that is true for the exponential and normal distributions too.

But there is a related concern that is harder to dismiss. In a Pareto distribution with $\alpha \leq 1$, the expectation is completely driven by the tail — by the possibility of durations orders of magnitude beyond any that have empirical support. And this is a general concern about distributions with infinite expectation: attempts to analyse them by their means (or even related techniques such as tail behaviour or stochastic dominance) suffer from extreme sensitivity to the behaviour of the tails far beyond the furthermost datapoint. This can be especially problematic if there is the possibility of a new, unmodelled, phenomenon cutting off the extreme values. The insensitivity to the tails of thin- and medium-tailed distributions makes them more robust to such possibilities.

These sensitivity concerns are smaller when just using the Pareto distribution to estimate the median time remaining, or related measures such as the 90th centile of time remaining. But caution should still be exercised if estimating extreme centiles (or starting with a very large $t_0$) since in many cases the Lindy effect may only apply within some bounded domain.

**Applications**

While human lifespans do not exhibit the Lindy effect, some animal lifespans do. Population ecologists study animal lifespans via *survivorship curves* (Pearl & Miner 1935; Deevey 1947). These are a rescaled version of the survival curve: survival is displayed on a logarithmic vertical axis and the time co-ordinate is normalised by the age at which exactly 1% (or some other small, fixed fraction) of the population remains (see *Figure 5*). The rescaling has the advantage that exponential decay of population (and thus a constant hazard rate) is shown as a straight line, making it easier to tell if the hazard rate is increasing or decreasing.

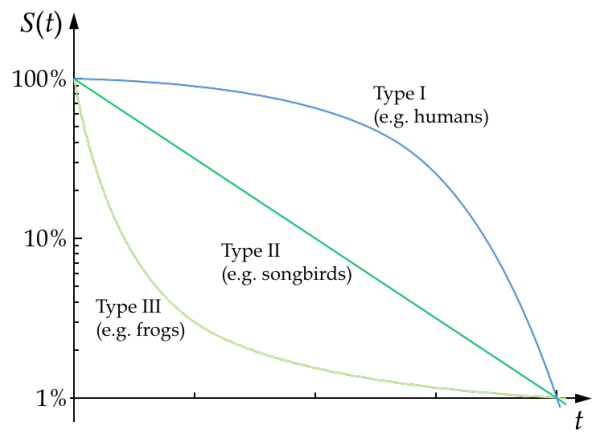

*Figure 5. Survivorship curves.* If we plot the logarithm of survival against (normalised) time we get a 'survivorship curve'. These are a popular tool in population ecology for categorising different patterns of mortality over time. Stylised curves for each of the three types are shown.

Animal species are categorised by whether the hazard rate is increasing (Type I), constant (Type II), or decreasing (Type III). Type III species include many that deploy the *r*-selection strategy for reproduction: putting their efforts into making a large quantity of offspring, rather than on shepherding each one through to reproductive age. Such species often have short generation times, small bodies, and the ability to widely disperse offspring. They include many insects, rodents, frogs, and grasses.

Because they have a decreasing hazard rate, Type III species exhibit the Lindy effect. They are thus a counterexample to Taleb's contention that human lifespans aren't subject to the (broad) Lindy effect because humans are perishable. Type III species are also perishable yet display the effect. Or at least they do over a substantial fraction of their lifespan. They may not be subject to the Lindy effect in the limit of high $t$ (where cellular damage and other consequences of aging may eventually win out and force the hazard rate up). But it will still be the case that for substantial domain of ages, median (and expected) remaining lifespan increases with age.

It may also be possible to observe a Lindy effect across a grouping of animals that includes many different species — even if no Type III species are included. For example, a set of different Type II species will typically have different constant hazard rates, and thus the joint population will exhibit a hazard rate that declines from the average of all the rates towards the lowest rate in the set, so could exhibit a Lindy effect.

This is part of a general tendency where mixing a sufficiently diverse set of homogenous populations can produce a Lindy effect, even when no individual population in the mixture does. While Taleb (2012) states that the Lindy effect applies at all levels of granularity (e.g. to every type of car as well as to cars in general), this need not be the case. For example, radioactive isotopes in general are Lindy, while no particular kind of isotope is.

There are also intimate connections between the Lindy effect and the practice of discounting in economics. Economic discounting involves valuing benefits less the further off in time they come (Ramsey 1928). There are several rationales including the fact that economic growth will make people richer over time (such that each extra dollar will be worth less to them) and uncertainty about whether the benefit stream (or its recipients) will still be around as time goes on.

The discount rate is often taken to be constant, leading to an exponential discounting of benefits over time (e.g. Koopmans 1960). But Martin Weitzman (1998) showed that if we account for our uncertainty about which constant rate to use (e.g. due to uncertainty over the future growth rate), then the effective discount rate can decline over time. While the underlying phenomenon is different, the mathematics of uncertain constant discount rates is identical to the mathematics of uncertain constant hazard rates that we saw earlier.

And there is an even closer connection between discounting and the Lindy effect. A stream of benefits is a process with an indefinite, uncertain lifespan. When exponential discounting is used to address this uncertainty, it effectively assumes the process suffers a constant hazard rate and so its future lifespan is independent of its current age. But as we've seen, uncertainty about the size of the hazard rate disrupts this exponential form and can produce a narrow Lindy effect. So can knowledge about the true shape of the survival curve for the particular process. Adapting economic practice to properly handle uncertain hazard rates via the resulting declining discount rates could lead to superior assessment of the costs and benefits of projects and policies.[16]

**Conclusions**

The Lindy effect is an important statistical phenomenon, appearing across many fields in many guises. We have explored definitions for both broad and narrow Lindy effects, identified the statistical distributions corresponding to them, and found a wide variety of generating mechanisms. These mechanisms include a vague Bayesian prior over the total lifespan, uncertainty about the hazard rate, a population with heterogenous hazard rates, and random fluctuations of robustness. We saw that while antifragility can produce a Lindy effect, so can a process of gaining increasing evidence about how fragile something is, or a process of winnowing out the most fragile members of the population. And we saw that there is an empirical form of the Lindy effect (where it shows up directly in observed lifespans, such as Type III species) as well as an epistemic form (where it shows up in our best estimates for the lifespans).

For this epistemic form, we saw that the Lindy estimate can be a good starting estimate when you have little information — perhaps even optimal if the current age is your *only* relevant information. And the Lindy estimate is also good in the common situation of uncertainty over the size of a slowly varying hazard rate. As more information comes in about the shape of the survival curve for the entity in question, the Lindy estimate may become less accurate than other ways of estimating. But it will always remain well calibrated.

**Funding**

This work was funded by a grant from Open Philanthropy.

**Acknowledgements**

I'd like to thank Finlay Moorhouse and Anders Sandberg for their helpful comments and discussions.

---

[16] It would also raise challenges. One reason for discounting was to allow valuations over the indefinite future while avoiding infinities. As the α = 1 Lindy effect has an infinite expectation, the challenges of comparing options with infinite valuations could return and would need to be dealt with in some other way.

**References**

Bailey, Robert T. (1997). 'Estimation from Zero-Failure Data.' *Risk Analysis*, 17(3), 375–80.

Carter, Brandon (1974). 'Large Number Coincidences and the Anthropic Principle in Cosmology'. *IAU Symposium 63: Confrontation of Cosmological Theories with Observational Data*, 291–298.

Carter, Brandon (1983). 'The anthropic principle and its implications for biological evolution'. *Philosophical Transactions of the Royal Society of London*, A310 (1512): 347–363. doi:10.1098/rsta.1983.0096.

Deevey, Edward S. Jr (1947). 'Life Tables for Natural Populations of Animals', *The Quarterly Review of Biology*, 22(4): 283–314.

Eliazar, Iddo (2017). 'Lindy's Law', *Physica A*, 486: 797–805, http://dx.doi.org/10.1016/j.physa.2017.05.077

Goldman, Albert (1964). 'Lindy's Law', *The New Republic*, June 13: 34–35.

Gott, J. Richard, III (1993). 'Implications of the Copernican principle for our future prospects'. *Nature*, 363 (6427): 315–319. doi:10.1038/363315a0.

Gott, J. Richard, III (1994). 'Future prospects discussed'. *Nature*, 368: 108. doi:10.1038/368108a0.

Gott, J. Richard, III. (1996). 'Our future in the universe', In ASP Conf. Ser. 88, *Clusters, Lensing and the Future of the Universe*, ed. V. Trimble & A. Reisenegger (San Francisco: ASP), 141.

Koopmans, Tjalling C. (1960). 'Stationary ordinal utility and impatience'. *Econometrica*, 28: 287-309.

Laplace, Pierre-Simon (1814). *Essai philosophique sur les probabilités*. Paris: Courcier.

Mandelbrot, Benoît (1982). *The Fractal Geometry of Nature*, W. H. Freeman and Company.

Miller Jr, Rupert G. (2011) *Survival analysis*, vol. 66. John Wiley & Sons.

Pearl, Raymond and John R. Miner. (1935). 'Experimental studies on the duration of life. XIV. The comparative mortality of certain lower organisms'. *Quart. Rev. Biol.*, 10: 60-79.

Quigley, John, and Matthew Revie. (2011). 'Estimating the Probability of Rare Events: Addressing Zero Failure Data.' *Risk Analysis*, 31(7), 1120–32.

Ramsey, Frank P. (1928). 'A Mathematical Theory of Saving.' *The Economic Journal*, 38(152), 543.

Taleb, Nassim Nicholas (2012). *Antifragile: Things that Gain from Disorder*, Random House.

Taleb, Nassim Nicholas (2021). 'Lindy as Distance from an Absorbing Barrier'. [Unpublished manuscript].

Tyson, Neil deGrasse, Michael A. Strauss, & Richard J. Gott III (2016). *Welcome to the Universe: An astrophysical tour*. Princeton University Press.

Weitzman, Martin. L. (1998). 'Why the Far-Distant Future Should Be Discounted at its Lowest Possible Rate.' *Journal of Environmental Economics and Management*, 36(3), 201–8.